\documentclass[preprint, aps,floats,floatfix,superscriptaddress]{revtex4}
\usepackage{graphicx} 
\usepackage{bm} 
\usepackage{epsfig}
\usepackage{pstricks}

\begin{document}

\title{Gravity and its Mysteries: Some Thoughts \& Speculations$^\dag$}

\author{A. Zee}
\affiliation{Department of Physics, University of California, Santa Barbara, CA 93106}
\affiliation{Kavli Institute for Theoretical Physics, University of California, Santa Barbara, CA 93106}

\begin{abstract}
I gave a rambling talk about gravity and its many mysteries at Chen-Ning Yang's 85th Birthday Celebration held in November 2007. I don't have any answers.\\

${^\dag}$Invited plenary talk, to appear in the Proceedings

\end{abstract}


\maketitle

It is an honor for me to be giving this talk on the occasion of Professor Chen-ning Yang's 85th birthday. Like many ethnic Chinese physicists of my generation, I was inspired to go into physics by accounts of the work of Lee and Yang on parity violation. When the organizers invited me, I clearly understood that I was not to talk about ``what I did last month" as is appropriate at a standard physics conference, but to give a broader perspective on some facet of theoretical physics. I am going to talk about how the mysteries of gravity have puzzled and fascinated me. Some of the following will reflect my own confusion and lack of understanding. I also confess to ignorance of entire chunks of the literature.

\section{The graviton knows about everything}
\label{introduction}

Gravity knows about everything, whatever its origin, luminous or dark, even the energy contained in fluctuating quantum fields.

As is well known, this leads us to one of the gravest puzzles of theoretical physics. Consider the Feynman diagram with the graviton coupling to a matter field (for example an electron field) loop. If we claim to understand the physics of the electron field up to an energy scale of M, then the graviton sees an energy density given schematically by $\Lambda \sim M^4+M^2 m_e^2 log(\frac{M}{m_e})+m_e^4 log(\frac{M}{m_e})+ \dots$. Just about any reasonable choice of $M$ leads to a humongous energy density!!! In fact, even if the first two terms were to be mysteriously deleted, there is still an energy density of order $m_e^4$, that is, an energy density corresponding to one electron mass in a volume the size of the Compton wavelength of the electron, filling all of space, which is clearly unacceptable.

Apparently, this disastrous prediction of quantum field theory has nothing to do with quantum gravity. Indeed, the quantum field theory we need for the matter field is merely free field theory: we are just adding up zero point energy of harmonic oscillators.

The cosmological constant paradox may be summarized as follows. In some suitable units, the cosmological constant was expected to have the value $\sim 10^{123}$. This was so huge that it was decreed to be equal to $=0$ identically, while the
measured value turned out to be $\sim 1$.  

I am presuming that the observed dark energy is the fabled cosmological constant. The evidence seems increasingly to favor this simplest of hypotheses. Even if this were not the case, much of the paradox still remains.

I define $\Lambda$ by writing the Einstein-Hilbert action as $\int d^4 x \sqrt{g} (\frac{1}{G}R+\Lambda)$. It is useful to define the mass scale of the cosmological constant according to 
$\Lambda \equiv {M_\Lambda}^4$. Since observationally the cosmological constant almost closes the universe we could write the Einstein-Friedmann equation $({\frac{1}{a}}{\frac{da}{dt}})^2=G\rho$ as $L_{\rm universe}^{-2}\sim G \Lambda$ with $L_{\rm universe}$ the size of the universe, say the Hubble radius. Let us define $M_U\equiv 1/L_{\rm universe}$ as some sort of Compton mass of the universe. Then we have $M_U^2 \sim {M_\Lambda}^4/ M_{\rm Pl}^2$ so that 
\begin{equation}
M_\Lambda \sim {\sqrt {M_{\rm Pl} M_U} }
 \end{equation}
With $M_{\rm Pl}\sim10^{19} ~Gev$ and $M_U\sim 2 \times 10^{-33} ~ev$ we find that $M_\Lambda \sim 4 \times 10^{-3} ~ev$.

Neutrino masses, while possibly quite different from family to family, appear to have generic values, very roughly, of order $10^{-3} ev$. Is this just a coincidence? In any case, there might be some physics we have yet to understand at a mass scale of $\sim 10^{-3} ev$.

Instead of thinking about the cosmological constant as an energy density we could regard it as a sort of ``curvature" by moving a left parenthesis and writing the Einstein-Hilbert action as $\int d^4 x \sqrt{g} \frac{1}{G}(R+\lambda)$. Then $\lambda$ has the dimension of an inverse square of a length, which we define as $l_{\Lambda}$. Again, observationally, we know that the two terms in the action have comparable weight, and hence the length scale associated with the cosmological constant is of the order of the size of the universe. In other words, $l_{\Lambda}=M_{\rm Pl}/{M_\Lambda^2}\sim 1/M_U \sim L_{\rm universe}$.

 Incidentally, while $\Lambda$ was decreed to be identically zero by theorists, it was never banished by observational cosmologists, who needed it to reconcile various discrepancies in the data (for example, a universe younger than the earth due to an erroneous value of the Hubble constant in the 1930s and the clustering of the redshift data of quasars in the 1960s.)

Gravity, knowing  about everything, is the only interaction sensitive to a shift of the Lagrangian by an additive constant. In classical physics, additive constants do not affect the equation of motion. In quantum mechanics, experiments typically measure only energy differences $\Delta E$ and not the energies themselves.  The Casimir effect measures the change in vacuum energy $\Delta E$ before and after the mirrors are introduced, not the vacuum energy itself (as is sometimes erroneously stated.) But gravity knows about the vacuum energy $\sum \frac{1}{2}\hbar \omega$. 

Is the zero point energy $\frac{1}{2} \hbar \omega$ real? I should think so, since it comes directly from the uncertainty principle. The textbook demonstration of reality is of course the liquidity of helium at zero temperature, but in fact, during the early days of quantum mechanics, many of the greats were skeptical. At the 1913 Solvay Congress Einstein declared that he did not believe in zero point energy, writing to Ehrenfest that the concept was ``dead as a door nail." Pauli also had his doubts, but the experiment $\gamma+H_2 \rightarrow H+H$ convinced him. He was apparently the first to worry about the gravitational effect of the zero point energy filling space. He used for $M$ the classical radius of the electron and concluded that the resulting universe ``could not even reach to the moon!" With the passage of time people found ``better" things to worry about and the issue was forgotten until Zel'dovich raised it again in the late sixties.

\section{The proton lifetime as an analogy}

I would like to argue by analogy: this is a time-honored tradition in physics, historically often helpful and suggestive. Let us try to think of a physical quantity once expected to be huge, later decreed to be zero, then measured to be small but not zero. What I came up with was an alternative history of proton decay. It didn't happen exactly this way in our civilization, but it could have easily happened in some other civilization somewhere else. Suppose that in the early 1950Õs, a bright young theorist decided to estimate the rate $\Gamma$ for the decay $p \rightarrow e^{+} + \pi^0$. He wrote down the effective Lagrangian ${\cal L} \sim f \pi e p$, and comparing with the pion nucleon interaction ${\cal L} \sim g \pi n p$ he ``naturally" expected $f \sim\alpha g$ with a factor  of $ \alpha $ thrown in for isospin violation. Obviously, this naturally expected rate $\Gamma$ came out way too large by many many orders of magnitude. The rate was decreed to be identically zero, by Wigner I think, and the decree came with some nice sounding words like ``baryon number conservation," in a typical example of 
proof by authority. Even though in our own world only an upper bound on $\Gamma$ exists, we could easily imagine that in our alternative world $\Gamma$  was later measured to be tiny compared to the  ÒnaturalÓ expectation, but definitely non-zero. 

We now know the resolution of this huge paradox. It did not come from thinking about a theory of proton decay, or what the Òright mechanismÓ for it might be, but from a ``totally unexpected" direction, namely baryon spectroscopy. 

The Lagrangian ${\cal L} \sim f \pi e p$ with scaling dimension
\begin{equation}
\frac{3}{2} + \frac{3}{2} + 1 = 4
 \end{equation}
got transmuted into
${\cal L}\sim \frac {1}{M^2} qqql$ with scaling dimension 
\begin{equation}
\frac{3}{2} + \frac{3}{2} +\frac{3}{2} +\frac{3}{2}  = 6.
 \end{equation} Here $M$ denotes the mass scale of the physics responsible for proton decay. Instead of the proton $p$ and the pion $\pi$ fields we write the quark field $q$, and $l$ is just a fancier way of writing $e$. Note that Lorentz invariance requires 4 fermion fields rather than 2.

Remarkably, this boost in scaling dimension from 4 to 6 is enough to solve an enormous paradox!!! The reason is that it appears in the exponential.  Thus, now $\Gamma$ is proportional to $(1/M^2)^2 = 1/M^4$. We also need the matrix element $<p | qqql |\pi e>$, which is set by low mass scale physics and so should be $\sim m_p$ the proton mass. Hence by ``high schoolÓ dimensional analysis, we obtain $\Gamma \sim(m_p/M)^4 m_p$, which for $M$ significantly larger than $m$ could account for the ``absurdly" small value of $\Gamma$!

We could even imagine that in this alternative civilization a bright young theorist could have argued that the long lifetime of the proton pointed to quarks.  

Do we learn something from this story?  Anything?  Nothing?

If this story is somehow relevant to the cosmological constant paradox, we might ask whether we could in some way promote ${\cal L} \sim \Lambda \sqrt{g}$ with scaling dimension $0$ to an operator ${\cal O}$ with dimension $p$ so that the effective Lagrangian for the energy density becomes ${\cal L} \sim \frac{1}{M^{p-4}} {\cal O}$ with $M$ some new high mass scale? We would then obtain $\Lambda \sim \frac{1}{M^{p-4}} <{\cal O}> = (m/M)^p M^4$ with $<{\cal O}> = m^p$ and $m$ some new low mass scale.

With $m$ small enough and $M$ large enough and $p$ big enough, we could get the suppression we want.

I certainly do not have a detailed theory of how this could happen. One question: How could we promote $\sqrt{g}$ without also promoting the Einstein-Hilbert term $\frac{1}{G} \sqrt{g} R$? Interestingly, the same question arises in my historical analogy and it might be instructive to watch how the question was resolved: $\pi e p$ was promoted to $qqql$ with dimension jumping from 4 to 6 but $\pi n p$ was changed to $qqA$ (here $A$ denotes the gluon field) with dimension remaining at 4.

\section{Could gravity be part of a larger structure?}

Could Einstein-Hilbert be replaced by something more fundamental which could lead to $\frac{1}{G} \sqrt{g} R$ effectively at low energy much as quantum chromodynamics leads to the Yukawa theory?

I am not necessarily suggesting here that the graviton is composite. Indeed, there is a seemingly convincing
ÒargumentÓ against the graviton being composite. Consider the same Feynman diagram mentioned earlier, but at the point where the graviton couples to the electron we insert a form factor with some energy scale. The trouble is that the momentum $q$ carried by the graviton is in what I would call the ``extreme ultra infrared" with $q \sim 1/L_{\rm cosmological}\sim 0$ where $L_{\rm cosmological}$ denotes a cosmological distance scale. In other words, the universe could care less if the graviton is composite at an energy scale of say $1 ~Tev$.

The alternative may be that gravitation is part of a larger structure (perhaps along the line I sketched in  
Phys. Rev. Lett. 55 (1985) 2379.) We now understand the electromagnetic field as part of a larger structure.  
Gerard 't Hooft has given an elegant expression for the Maxwell field ${\cal F}_{\mu\nu}$ in terms of the Yang-Mills field ${F^a}_{\mu\nu}$. Is there an analog for gravity? The statement that the electromagnetic field is part of a larger structure, even if the structure is not seen at low energies, does lead to physical consequences. Thus, electric charge is quantized if the larger structure is a grand unified theory based on a simple group, and we understand why $Q_{\rm electron}=-Q_{\rm proton}$ exactly, a fact of cosmological significance.

Alternatively, we could argue that quantum field theory such as quantum electrodynamics only makes sense when formulated on a lattice, and then the electromagnetic $U(1)$ is ``necessarily" compact which leads to charge quantization. In either case, what we see and know is part of a larger structure. 

The question, stated in the format of an IQ test question, is then ``What is to gravity as Yang-Mills is to electromagnetism?"

\section{The horizon}

Many have made careers out of worrying about quantum gravity. But classical gravity is already plenty puzzling. When we first studied physics, we were told that physics should be local, that something happening here could only affect something happening nearby, and for a physical effect to propagate across spacetime a field is needed. (The mysteries of quantum mechanics have however also led to entanglement and the Aharonov-Bohm effect.) Already in classical, non-quantum, gravity we have black holes, and the horizon around a black hole is a strikingly non-local concept. ÒNothing happensÓ locally. Observers falling in do not notice anything. The hand-wringing over the horizon only affects the mythical observer stationed way off at infinity. To me that is a basic puzzle of physics. In technical terms, the Riemann curvature is nice and smooth at the horizon and could be made arbitrarily small for massive black holes. But somehow the other fields know about the metric $g_{\mu\nu}$ directly, not Riemann curvature. For a nice pedagogical treatment of how the horizon appears as a black hole forms, see the not terribly well-known work by R. Adler, J. Bjorken, P.S. Chen and J. S. Liu.

The horizon is an inherently non-local concept. But confusingly, while we cannot perform local measurements to detect the presence of a horizon directly, we could do so indirectly. By measuring  whether light rays tend to converge or diverge, we could detect the presence of a ``trapped surface" (or apparent horizon.) A sequence of highly plausible theorems (each of which nevertheless involves some technical assumptions) by Penrose, Ellis, and others, combined with the unproven cosmic censorship conjecture, states that the presence of a trapped surface implies the presence of a horizon. 

More physically, the horizon is non-local in the following sense. By drawing a Penrose diagram we can see that we could be sitting peacefully with an incoming shell of matter far away threatening to form a black hole soon and we might be inside the horizon even before the black hole forms. 

In the standard Schwarzschild coordinates, $g_{00}=0$ and $g_{rr}=\infty$ at the horizon. Time and space then exchange roles. It would appear that to have a proper formulation of quantum mechanics and quantum field theory we need to have a well-behaved time variable to evolve unitarily with. As is well known, there are textbook formulations of quantum field theory in curved spacetime and standard treatments lead to Hawking radiation. Are these treatments correct? Is there a modification of Einstein's theory such that metric singularities such as $g_{00}=0$ and $g_{rr}=\infty$ are somehow forbidden. Of course, any student  knows that  these are but artifacts due to a poor coordinate choice. We could transform to coordinates in which $g_{00}$ and $g_{rr}$ are perfectly well behaved at the horizon, for example the Kruskal-Szekeres coordinates.

Historically, the horizon was a source of great confusion and Kruskal's contribution cannot be overestimated. For example, on page 203 of Bergmann's standard text ``Introduction to the Theory of Relativity (with a foreword by A. Einstein)" he quoted Robertson as concluding that ``at least part of the singular character" of the metric at $r=2GM$ must be attributed to the choice of coordinates. Curiously, people at the time did not follow the modern expedient of simply noting the smoothness of the Riemann curvature tensor, which Schwarzschild himself, at the very least, must have calculated. (Bergmann then went on and cited Einstein's 1939 work showing that in a toy model of a spherical cluster of non-interacting particles the Schwarzschild singularity could not form.) 

Could we possibly modify general relativity so as to avoid having an horizon? Once again, apparently not because black hole is a low energy phenomenon. Naively, we might also think addition of local terms would not remove a non-local phenomenon like a horizon.

When we turn on quantum mechanics the black hole emits Hawking radiation and eventually disappears, so that the horizon is not only not local in space, but also not local in time.

Quantum field theory in curved spacetime is a well developed subject
and leads to Hawking radiation for example,  but again I still have
lingering doubts. In calculating a loop diagram for some quantity, say the
electron's magnetic moment, at the horizon, are there subtleties
involving virtual particles propagating inside the horizon and then
out again? Presumably it is okay over a distance scale of the order the Compton
wavelength of the particles involved. More generally, in doing a quantum
gravity path integral sum over all gravitational field configurations,
are we to include configurations containing black holes or not? I
imagine that there are experts walking around who are sure of the
answers to these questions and more.

\section{The gravitational field is not just another field}

According to an apparently appealing philosophy due to many eminent physicists ($\dots$, Gupta, Kraichnan, Feynman, Weinberg, Boulware, Deser, $\dots$), we should regard the gravitational field as just another field. As Feynman showed in his posthumous book on gravity, we could pretend that we never heard of general relativity and Riemannian geometry, and simply develop the field theory of a massless spin 2 particle called the graviton.  The program worked:  general relativity and Riemannian geometry emerge from playing with Feynman diagrams, but most people hate this anti-geometric approach. (By the way, Kraichnan did his work as an 18-year old undergraduate at MIT. According to his recollection, Einstein was appalled by this
approach. Partly as a result, Kraichnan  delayed publication for eight
years and so ended up publishing after Gupta. Feynman was apparently
unaware of the work of Kraichnan and Gupta.)

Nevertheless, this view is somehow fundamentally wrong. The way I like to put it is that we are in some avant-garde theater. Unique among the actors, the graviton is not just another actor on the stage.  The actor is himself the stage.  It provides the arena in which the other fields work and play.

The founders of quantum field theory wrote profound equations such as
\begin{equation}
A_\mu = 0 + A_\mu 
\end{equation}
and 
\begin{equation}
\varphi = 0 + \varphi   
\end{equation}
Fields execute quantum fluctuation around vanishing classical values. But then physicists became more sophisticated in the  1960Õs and wrote fancier equations like
 \begin{equation}
\varphi = v + h   
\end{equation}
with $v=<\varphi>$.
A great deal of money has been, and is being spent, to see if this idea is correct. 

The basic equation for the graviton field has the same form
 \begin{equation}
g_{\mu\nu} = \eta_{\mu\nu} + h_{\mu\nu}    
\end{equation}
This naturally suggests that $ \eta_{\mu\nu} = \langle g_{\mu\nu} \rangle$ and perhaps some sort of spontaneous symmetry breaking. But  gravity exhibits a  fundamentally new feature: $g_{\mu\nu}$ is a matrix, and hence has a signature. Large fluctuations of $h_{\mu\nu} $ could change the signature of $g_{\mu\nu} $ and there could be regions with two ÒtimesÓ. An obvious thing to write down would be a potential for  $g_{\mu\nu} $ (which breaks general coordinate invariance) of the form $V(g)=\lambda (g_{\mu\nu}- \eta_{\mu\nu})^2$, or more generally a potential with a deep well pinning $g_{\mu\nu}$ to values close to $\eta_{\mu\nu}$. This induces a graviton mass of order $m_g^2 \sim \lambda M_{Pl}^2$ so that $\lambda^{\frac{1}{2}}$ is given by the ratio of the largest mass  and possibly the smallest mass known to physics.

This line of thought raises the possibility that the potential $V$ might have minima elsewhere. Perhaps there is a phase with $g_{\mu\nu}=0$. That could be the ultimate terrorist plot, to unleash a $g_{\mu\nu}=0$ bomb that would annihilate spacetime in the victimized country.

An intriguing idea is that of emergent gravity developed by X.G. Wen and others. This line of development emerged from the  days of the chiral spin liquid, in which gauge fields readily emerged from systems that consist solely of electron spins. (See for example, A. Zee, in the M.A. B. B\'eg Memorial Volume 1990 edited by A. Ali and P. Hoodbhoy.) Besides gravity, fermions also puzzle me. (Jordan's manuscript languished in Born's pocket for a whole year.)  Sometimes I feel that the world ought to contain only bose fields. Perhaps half integral spin could also be emergent. (See for example, A. Zee in ``Quantum Coherence: 30 years of Aharonov-Bohm Effect" 1989 edited by J. Anandan.) I am also intrigued by the effect discovered by 't Hooft et al, that binding a boson to magnetic monopole produces a fermion.

\section{Unimodular gravity}

The notion of unimodular gravity goes back to Einstein in some sense, and was developed later by Anderson, Finkelstein, van der Bij, van Dam, Y.J. Ng, Wilczek, Zee, Dolgov, Weinberg, and many others. Suppose $g \equiv \det g_{\mu\nu}$  is fixed to be equal to 1, then the cosmological constant term in the action $S=\int {d^4x} {\sqrt{g}} \Lambda+....$ becomes impotent, and hence irrelevant. But in fact, it comes back. Since the constraint $\delta \det g_{\mu\nu}=0$ is equivalent to $g^{\mu\nu} \delta g_{\mu\nu}=0$ we only get the traceless part of Einstein's equation
\begin{equation}
R_{\mu\nu} - {\frac{1}{4}}g_{\mu\nu}R = T_{\mu\nu} -{ \frac{1}{4}}g_{\mu\nu}T
 \end{equation}
Writing $- {\frac{1}{4}}$ on the left hand side as $- \frac{1}{2}+ \frac{1}{4}$ and taking the covariant derivative, we obtain $\partial_\mu R=-\partial_\mu T$, which could be solved to give $R=-T+C$. The integration constant $C$ reappears as the cosmological constant when this equation is inserted back into the traceless part of Einstein's equation. 

Thus, unimodular gravity does not solve the problem but makes some people ``feel more comfortable" because in theoretical physics, supposedly, we have the license to set integration constants to whatever we want.

\section{Equivalence principle}

Let us go back to the Feynman diagram described at the beginning of this article, with the graviton coupling to a matter field, say the electron field, loop. Ultimately, it is this graph that causes all our hand-wringing over the cosmological constant. Suppose one were to work long and hard and come up with a ÒruleÓ or ÒtheoryÓ that ÒcleverlyÓ deletes this graph, thus solving the cosmological constant paradox. As emphasized by J. Polchinski, any such  ÒruleÓ or ÒtheoryÓ would always be doomed to fail because of the equivalence principle.  

The argument is as follows. Connect the graph by some photon lines to the propagator of some atomic nucleus, say aluminum or iron. This graph thus contributes to the gravitational mass of the nucleus. On the other hand, consider the same graph with the atomic nucleus but with the graviton removed, a graph that presumably has nothing to do with gravity. But this graph contributes to the inertial mass of the nucleus. Thus, with the enormous accuracy to which the equivalence principle has been tested, we already know that the graph with the graviton attached could not be deleted. But we are claiming that, in order to resolve the cosmological constant paradox, we have some ``rule" to delete this graph. 

The trouble is once again that physics as we understand it should be local: at the point the graviton couples to the electron, how could the graviton ``know"  what the electron loop is going to do? It could not know whether the electron is just going to loop back upon itself, or that before looping back, the electron is ``planning" to emit two photons which subsequently will be absorbed  by a nucleus.

\section{The extreme ultra infrared}

The local nature of Feynman diagrams, plus the constraint from the experimental verification of the equivalence principle, make it difficult to imagine how any ``rule" could be invented to delete one Feynman diagram and not another. Perhaps one loophole is offered by the phrase ``nothing to do with gravity"; perhaps even a graph without the graviton is subject to the requirements of some ultimate theory of gravity.  

Another way out is suggested by the fact that, upon closer inspection, we see that there is evidently  a huge difference  between the graph responsible for the cosmological constant paradox and the same graph attached by two photon lines to a nucleus. The momentum carried by the graviton, called it $q$, has a value $q \sim 1/L_{\rm{cosmological}}$ in the former, but a vastly larger value  $q \sim 1/L_ {\rm{laboratory}}$ or $q \sim 1/L_ {\rm{terrestrial}}$ in the latter.

In particle physics we always profess ignorance about physics at high energies, about the ultraviolet regime, but truth be told, we know almost nothing about the ``extreme ultra infrared." Thus, we could always modify the left hand side of Einstein's equation by acting with some operator $f(L^2 D^2)$ where $D$ denotes the covariant derivative and $L$ is some cosmological length scale. The left hand side is effectively multiplied by $f(L^2/L^2_{\rm phenomenon})$ where $L_{\rm phenomenon}$ denotes the length scale of the phenomenon under study. All we require in order to distinguish between the two graphs is for $f$ to have the properties $f(\sim \infty) = 1$ and $f(\sim 0) = 0$. Needless to say, such a momentum dependent function implies that the theory is highly non-local.

\section{The universe is secretly acausal but only the universe knows about it}

One realization of this sort of idea is due to Arkani-Hamed et al (2002) who proposed modifying Einstein's equation to
\begin{equation}
M^2_{\rm Pl} (R_{\mu\nu} - {\frac{1}{2}}g_{\mu\nu}R) -{ \frac{1}{4}}{\bar M^2} g_{\mu\nu} {\bar R}=T_{\mu\nu}
 \end{equation}
 where $ {\bar R}$ denotes the space-time averaged scalar curvature  ${\bar R} \equiv \frac {\int d^4 x \sqrt{g}R}{ \int d^4x\sqrt{g} }$.
This equation is manifestly non-local and acausal:  physics now depends on physics in the far future. But by construction the modification to Einstein's equation takes effect only if the future is de Sitter with constant scalar curvature determined by the cosmological constant ${\bar R} = -\frac{4\Lambda}{M_{\rm Pl}^2 + {\bar M^2} }$. To account for observation, the new mass scale ${\bar M}$  has to be  huge, taking values ranging from $\sim10^{48} ~Gev$ to $\sim10^{80} ~Gev$ depending on the assumed value of the cosmological constant one wishes to ``neutralize." Unhappily, another enormous mass scale has to be introduced into physics.

In this approach, the modification is clearly designed not to matter for any situation other than cosmological. For the solar system for example, ${\bar R}$ would come out to be practically zero. The universe is secretly acausal but only the universe knows about it! I must say that in recent years, theoretical physicists have become increasingly adept at hiding new physics from experimentalists.

Arkani-Hamed at al argued that any mechanism to ``neutralize" the cosmological constant must be acausal: when a vacuum energy density ``turns on", the alleged mechanism must ``wait" for a cosmological time period to ``find out" if the energy density is indeed a cosmological constant. I am very much troubled by the thought that physics may be ultimately non-local, but the argument appears to be plausible.

\section{Induced gravity}

At one time induced gravity appeared to offer a way out of our problems with gravity and thus enjoyed a following. Consider the path integral
\begin{equation}
\int d\phi d\psi dA e^{iS(g,\phi,\psi, A)}= e^{i\int d^4x {\sqrt g}(\Lambda + R/G+\dots)}
 \end{equation}
There is no question that integration over the matter fields $\phi$, $\psi$, and $A$ would generate the Einstein-Hilbert term. The difficulty is that $\Lambda$ comes out naturally large, but this is of course just the cosmological constant problem again.

One fundamental question is whether we need to integrate next over $ {\cal D}g$. If not, that is, if we do not integrate over the metric, then the classical equation of motion of the gravitational field would not emerge automatically as Planck's constant approaches zero but has to be imposed by hand.

This leads us to the perennial question of whether gravity has to be quantized. If not, as was first proposed by M\o ller  (1962) and Rosenfeld (1963), then we have the equation
\begin{equation}
R_{\mu\nu}-{\frac{1}{2}}g_{\mu\nu}R=\langle T_{\mu\nu} \rangle
\end{equation}
Once again, this produces a huge cosmological constant on the right hand side. But let us leave that aside. The objection to this equation is that it violates the uncertainty principle. If gravity is not quantized, then it acts as a classical probe, and we could use a massive ball attached to a torsion balance to measure the position and momentum of a passing electron. In 1981 Page and Geiliker experimentally demonstrated the difficulty one runs into. Consider a Cavendish experiment in which the heavy ball is moved from one position ``here" to another position ``there" as determined by some radioactive decay. This amounts to a Schr\"odinger cat experiment with the quantum state in the preceding equation given by $|\rangle  ={\frac{1}{\sqrt 2}}(|here \rangle +|there \rangle)$ The torsion pendulum would then point to a ``phantom ball" situated half-way between here and there.

There are those (for example Dyson) who would raise the question of whether gravity has to be quantized on phenomenological grounds, since no conceivable experiment could detect a single graviton.

\section{Ever more speculative ways out}

Over the years, many physicists have had many (``crazy") thoughts about gravity. I listed some of them in a talk almost a quarter of a century ago on an occasion similar to this one, dedicated to Paul Dirac. One possibility, considered highly speculative at the time, was to entertain a decaying cosmological constant $\frac{d\Lambda}{dt}\neq 0$, but these days, with a multitude of scalar fields around, this possibility would be considered commonplace rather than outrageous. (See ``High Energy Physics in Honor of P. A. M. Dirac in His 80th Year", ed. by S. Mintz et al 1983.)

The cosmological constant paradox suggests to some people that we might have to break free of local field theory entirely. This line of thought led Steve Hsu and I to propose adding terms not of the form $\int d^4x (\cdots)$ to the action, in a vaguely ``Landau-Ginzburg" sort of approach to the action. We obtained
\begin{equation}
M_\Lambda \sim {\sqrt {M_{\rm Pl} M_U} }
\end{equation}
where $M_U$ was defined earlier as the Compton mass of the universe. This relation, regardless of how shakily it is derived, has the pleasing form of giving the mass scale of the cosmological constant (or dark energy) $M_\Lambda$ as the geometric mean of perhaps the largest and smallest mass scales in physics $M_{\rm Pl}$ and $M_U$. As explained earlier, it goes back to Einstein since it amounts to the statement that the observed dark energy is just about enough to close the universe.

\section{Reversal of fortune}

We have witnessed a remarkable shift in attitude towards quantum field theory over the last 30 years. An operator in the action is classified according to whether its mass dimension is $<4$, $=4$, or $>4$, operators known respectively as ``Super-Renormalizable," ``Renormalizable," and ``Non-Renormalizable." Textbooks taught that super-renormalizable interactions are nice, renormalizable interactions are what we want, while non-renormalizable interactions should fill us with fear and loathing.

This traditional doctrine was replaced by a new attitude which regards quantum field theory as a low energy effective theory. In an astonishing reversal of fortune, the non-renormalizable terms are now welcomed and well-liked as terms that are inevitably here with us. They are regarded as innocuous since they are suppressed by powers of some higher mass scale $\frac{1}{M^p}$, while the renormalizable terms are uniquely fixed by the gauge principle etc. In contrast, our ``friends" the super-renormalizable terms are now regarded as nasty guys.

Since these nasty guys have nominal mass dimension $<4$, there are fortunately only a finite number of them. They represent the challenges confronting fundamental physics today, and are in turn known as the Higgs mass term, the Einstein-Hilbert term, and the cosmological constant term. The Higgs mass term has dimension 2. The Einstein-Hilbert term has nominal dimension 2 which after rescaling by the Planck mass becomes dimension $4 + 5 + 6 + \dots$. The cosmological constant term has nominal dimension 0 which after rescaling becomes dimension $0 + 1 + 2 +\dots$.

Perhaps there is something seriously wrong with this picture.

Our understanding of physics is based on this notion of effective field theory, to which all we know could be reduced. Yet there are many questions, many doubts, but no clear answer. Field theory itself, and Einstein gravity as an effective field theory, could fail at truly long distances. Ultraviolet regularization has been understood for long time, but as I have said, not the extreme infrared. 

Quantum field theory is very much based on the momentum-distance relation, also known as the uncertainty relation, as expressed in the Fourier relation $e^\frac{ipx}{\hbar} $. This connection could fail and be modified. (Indeed, this is what happens in string theory.) From the discussion of the cosmological constant paradox it is clear that some kind of connection between ultraviolet and infrared (such as that offered by the anthropic principle: $\Lambda$ is ultraviolet while we humans are infrared) is needed.

Black hole offers a well known ``violation" of the standard momentum-distance relation: the more massive the black hole, the larger its size  $R=GM$. Clearly, the exception is due to the existence of a fundamental mass scale $l_{\rm Pl}=\frac{1}{\sqrt{G}}$.

Another possibility is the breakdown of quantum mechanics when the splitting between energy levels $\Delta E$ is less than the inverse of some cosmological time scale, such as the age of universe.

Meanwhile, Bern, Kosower, and many others, using the twistor formalism, have discovered amazing cancellations and simplifications in complicated Feynman diagram calculations. I find particularly intriguing the hint from the explicit calculation performed by Bern et al that amplitudes in Einstein gravity could be regarded as the square, or sum of squares, of appropriately ``color stripped" amplitudes in Yang-Mills theory, a hidden relation originally suggested in a string theory context by Kawai, Lewellen, and Tye. Recent work by Arkani-Hamed and others give tantalizing evidence that superficially more complicated theory like Einstein gravity and Yang-Mills theory may have better ultraviolet behavior than a simple scalar field theory. 

I have always been bothered by the liberal and indiscriminate use of scalar fields in particle theory and cosmology. Quantum field theory textbooks start with scalar fields precisely because they are ``without qualities". If Nature wanted to show us an elementary scalar field, wouldn't she have shown us one long ago? We have encountered elementary spin $1$ fields, an elementary spin $2$ field, and in a mysterious twist, even elementary spin $1/2$ fields. We know about meson fields, but they are clearly composite. When and if the Higgs field is discovered, an interesting questions might be whether or not it could be regarded as composite. I have speculated elsewhere that perhaps quantum field theory somehow forbids elementary scalar fields. In a
new formulation of quantum field theory, and one appears to be suggested by recent work, might elementary scalar fields not be allowed?

Note that in our historical analogy, when the pseudoscalar $\pi$ field was banished in favor of two quark fields the scaling dimension of the relevant operator goes up by $2$ and physicists have one less ``naturalness" paradox to contend with.

\section{Hierarchy problems}

It seems to me that the discovery of a small non-vanishing cosmological constant may have liberated us from having to worry about the various hierarchy problems of particle physics. The small cosmological constant, if indeed a cosmological constant, would be a living exception to the 't Hooft ``naturalness doctrine" regarding the occurrence of small dimensionless numbers in physics. In practical terms, one of the arguments in favor of the rather unlikely and contrived idea of low energy supersymmetry might have evaporated.

\section{The coincidence problem}

No discussion of the cosmological constant paradox is complete without mentioning the cosmic coincidence problem. The energy density $\rho$ in
matter varies with the scale factor $a$ of the expanding universe like $1/a^3$, while the energy density in
curvature varies like $1/a^2$, and the energy density in
the cosmological constant varies like $1/a^0$. It is remarkable that they are comparable now. Why now?!?

The only plausible ``explanation" is the ``anthropic lack of principle." In some sense, the smallness of $\Lambda$ was predicted by Weinberg using a very weak version of the anthropic principle. This very weak version of the anthropic principle should be acceptable to most theoretical physicists: it merely correlates two observations, namely that galaxies formed and the smallness of $\Lambda$.

\section{Closing remarks}

I was recently reading about the history of special relativity. Young Einstein was able to accomplish what Lorentz and Poincar\'e were not able  to accomplish, even though the two established giants had most of it worked out, at least mathematically. After all, Lorentz had the Lorentz transformation in all its glory. The two older physicists were not able to abandon the perfectly sensible notion that if there is a wave $something$ must be waving. So they had the ether as a dynamical variable. Einstein simply trashed the ether and asserted that $nothing$ could also wave.

Nowadays, any student is able to accept, without blinking twice, that an electromagnetic wave consists of $A_{\mu}$ waving, yes, just a mathematical symbol known as a field waving. Of course, there are energy and momentum densities associated with the wave, and so it is real in that sense. But what is a field? After spending years writing a textbook on quantum field theory, I could understand a field as only something that does what a  field does. No more, no less.

To move forward, physics had to abandon an apparently ironclad piece of commonsense that ``where there is a wave $something$ must be waving." I would not be at all surprised if it turns out that to move forward, we have to abandon an equally ironclad piece of commonsense. I leave it  to the reader to  identify that piece.

We conclude with a rather dark motto about dark energy I learned after giving a related talk in Bologna: "Per obscura ad obscuriora."\\

{\it Acknowledgments--} Over the decades I have benefited from conversations about gravity with many colleagues far too numerous to name here, starting with John Wheeler who guided me through my very first research project. During this past year leading up to this talk, and in preparing for this talk, I was enlightened on various occasions by Nima Arkani-Hamed, Steve Hsu, Joe Polchinski, Xiao-gang Wen, and Frank Wilczek. I thank Joe Polchinski and Rafael Porto for reading this manuscript and Ted Jacobson for helpful comments. I am also grateful to Richard Neher and Rafael Porto for technical help in preparing this article.

Some versions of this talk were also given at the Lorentz Institute, Leiden, the Netherlands, the National Taiwan University, Taipei, Republic of China, and the Institute for Theoretical Physics, Sao Paulo, Brasil.

I am  supported in part by NSF under Grant No. 04-56556.\\

\end{document}